\documentclass[twoside]{ilcws10}
\usepackage[latin1]{inputenc}
\usepackage[dvips]{graphicx,epsfig,color}
\usepackage{wrapfig,rotating}
\usepackage{amssymb,amsmath,array}

\begin{document}
\title{
%%%%   Paper title goes here  %%%%%%%%%%%%%%
Model independent WIMP Searches in full Simulation of the ILD Detector} %% 
%***********************************************************************
% AUTHORS INFORMATION AREA
%***********************************************************************
\author{Christoph Bartels$^{1,2}$ and Jenny List$^1$
% Optional short acknowledgment: remove next line if non-needed
%\thanks{This is an optional funding source acknowledgment.}
% DO NOT MODIFY THE FOLLOWING '\vspace' ARGUMENT
\vspace{.3cm}\\
% Addresses and institutions (remove "1- " in case of a single institution)
1- DESY  \\
Notkestrasse 85, 22607 Hamburg - Germany
%% Remove the next three lines in case of a single institution
\vspace{.1cm}\\
2- University of Hamburg - Institut f\"ur Experimentalphysik \\
Luruper Chaussee 149, 22761 Hamburg - Germany\\
}
%%***********************************************************************
% END OF AUTHORS INFORMATION AREA
%***********************************************************************

\maketitle

\begin{abstract}
In this study the ILC's capabilities for detecting WIMPs and measure
their properties are investigated. The signal events 
are detected by associated production
of Initial State Radiation~(ISR). A model independent
formulation of the signal cross section is used. The cross section is normalized
by inference from the observed abundance of cosmological Dark Matter~(DM).
The study is performed in full simulation of the {\tt ILD00} detector
model. The prospects of determining the WIMP parameters
individually and simultaneously are presented.
\end{abstract}

\section{Dark Matter and WIMPs}\label{Sec:2}
\begin{wrapfigure}{r}{0.5\columnwidth}
\centerline{\includegraphics[width=0.45\columnwidth]{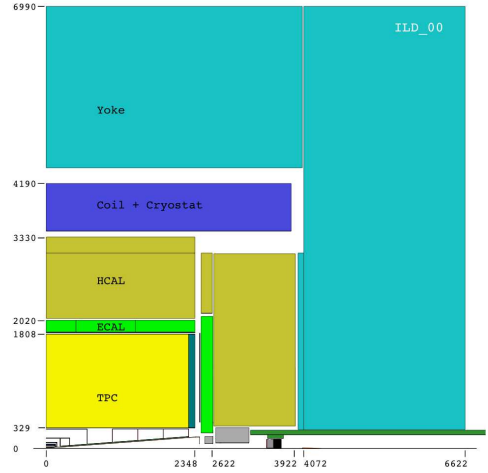}}
\caption{Quadrant of the ILD detector. Starting from the origin of
the coordinate system the Time Projection Chamber~(yellow) 
and the calorimetric system of ECAL~(green) and HCAL~(ocher) are shown.}
\label{Fig:Quad}
\end{wrapfigure}
New physics is required from both cosmology and the known shortcomings of the 
Standard Model~(SM).
The observed non-baryonic Cold Dark Matter~(CDM) density in the universe can
be accounted for by introduction of a new neutral particle,
stable by virtue of a new conserved quantum number. Given
interactions of weak strength and masses in the
range of 0.1--1~TeV these generic WIMPs naturally
give the observed DM density as a thermal relic of
the expansion of the early universe. Several
extensions of the SM include such a WIMP candidate,
e.~g.~SUSY with conserved R-Parity, where the LSP often is
either the lightest neutralino ${\tilde \chi}^0_1$ or the
gravitino ${\tilde G}$. 
Direct WIMP pair production $e^+e^-\rightarrow\chi\chi$ should be 
possible at the International Linear Collider~(ILC)~\cite{Brau:2007zza} for WIMP masses of
less than half the CMS energy, $M_\chi \le \sqrt{s}/2$. However, since WIMP
interactions are only of weak strength, such events would
be unobservable as long as no other detectable particle 
is associated with the pair production process. 
ISR provides such a detectable particle.
\begin{figure}[ht]
\setlength{\unitlength}{1cm}
\begin{picture}(13.0,3.0)
\put(0.0,0.0){\epsfig{file=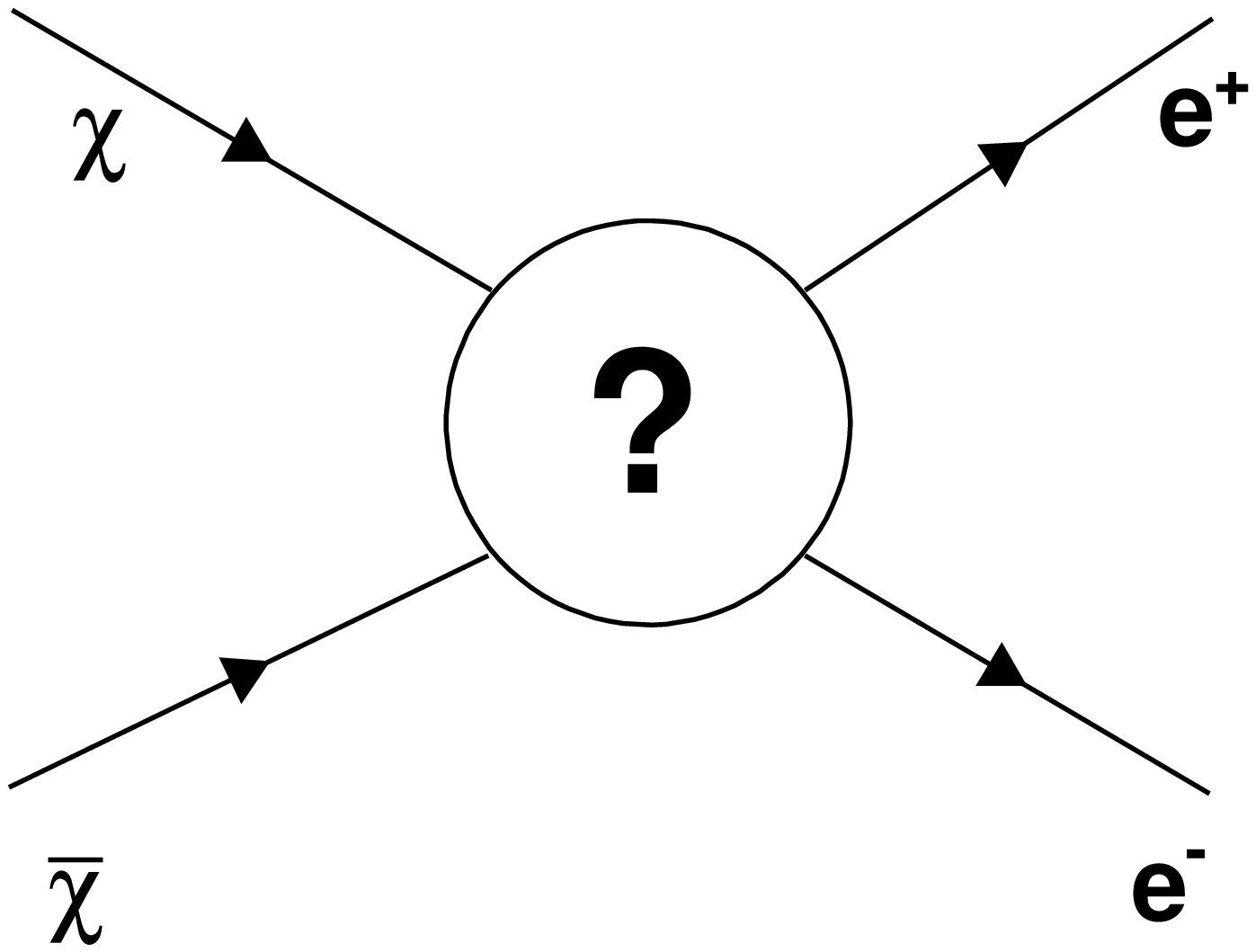,scale=0.25}}
\put(4.5,0.0){\epsfig{file=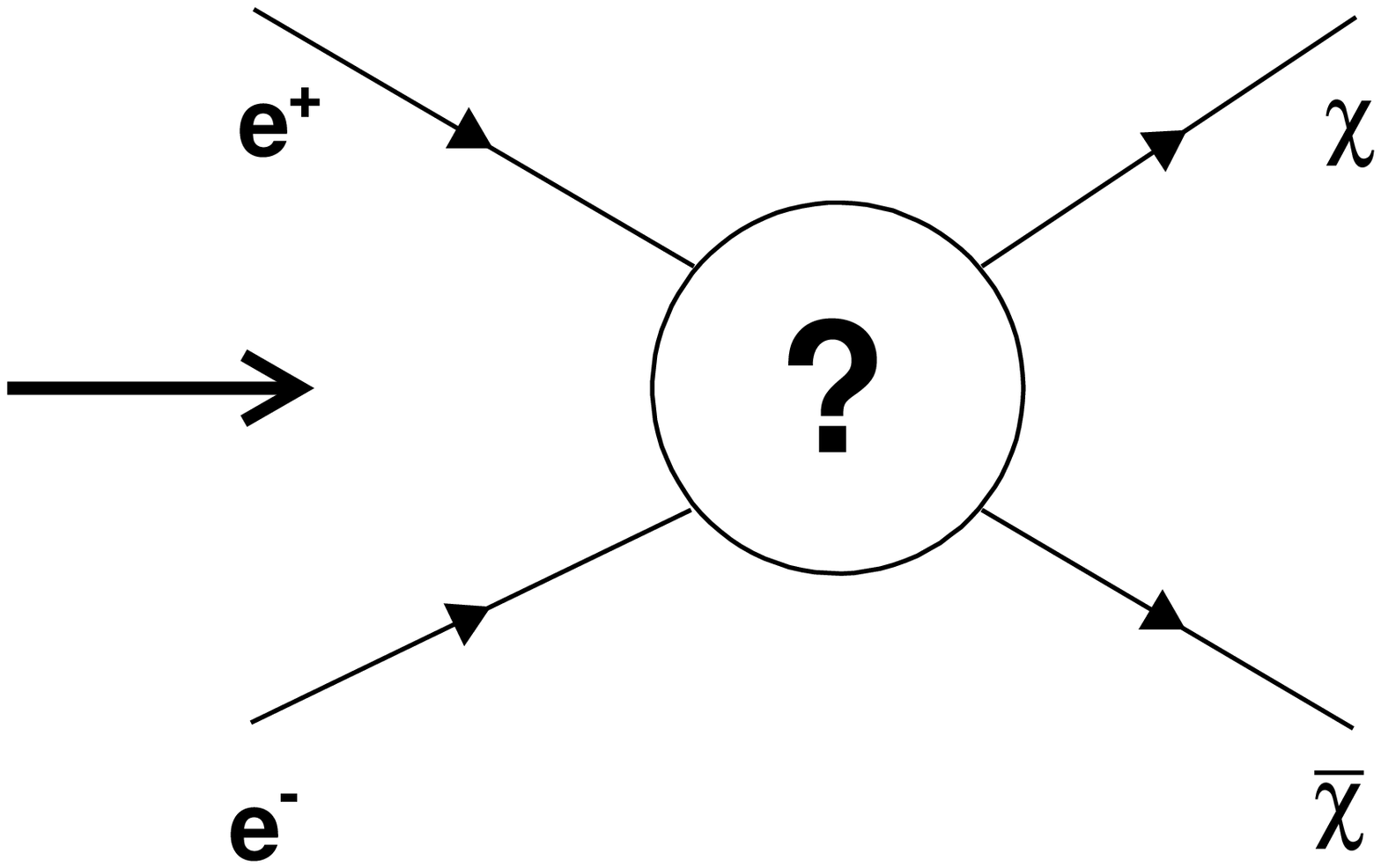,scale=0.25}}
\put(9.0,0.0){\epsfig{file=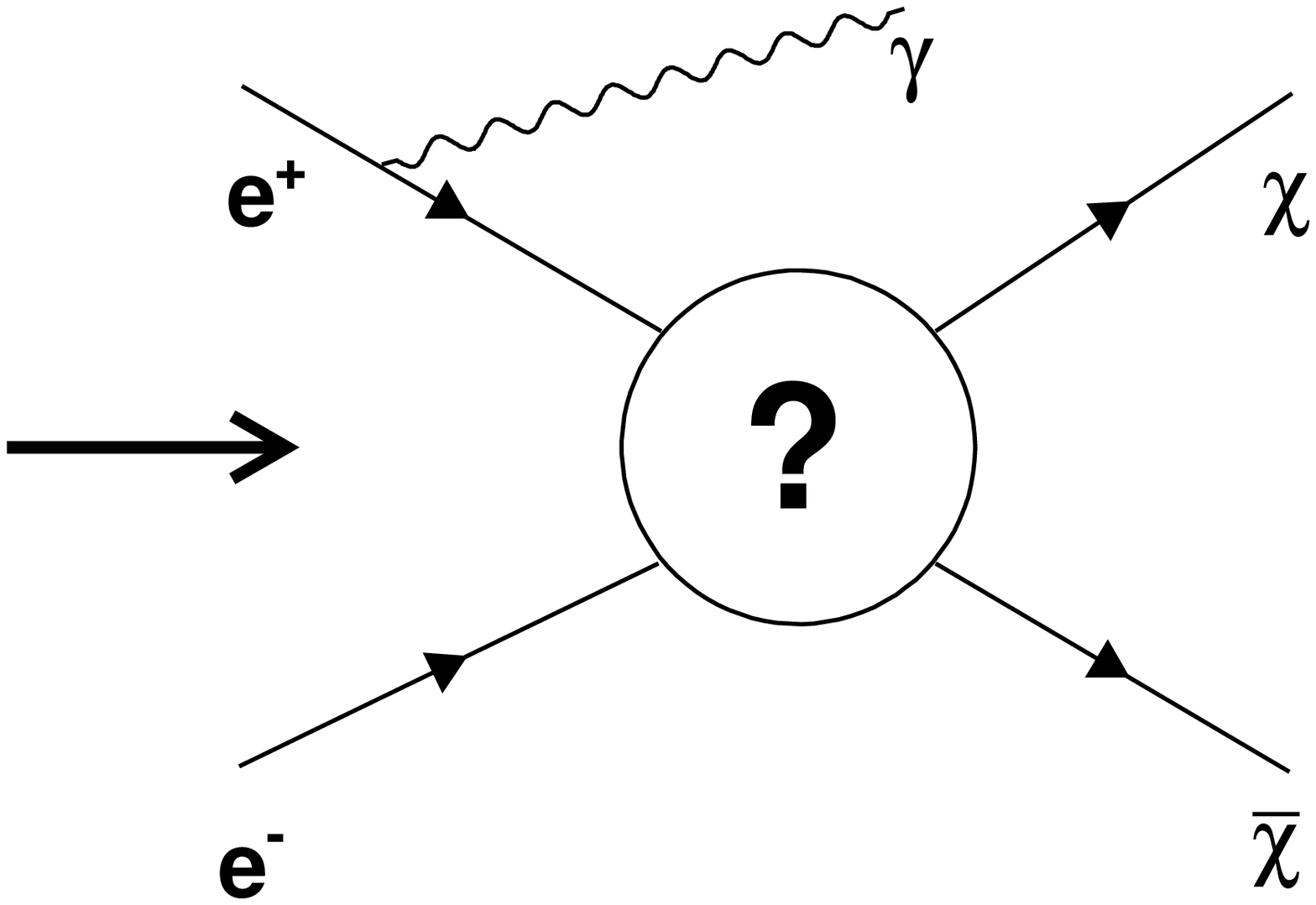,scale=0.25}}
\put(0.0,0.0){(a)}
\put(4.5,0.0){(b)} 
\put(9.0,0.0){(c)}
\end{picture}
\caption{Schematic derivation of the $e^+e^-\rightarrow\chi\chi\gamma$ production cross section. 
Except for some generic assumptions on the WIMP nature no presumptions on the
interactions involved are imposed, indicated with the question marks.}\label{Fig:Birk}
\end{figure}
A single photon radiated off the initial state delivers 
a clean signal, while the properties of the photon
spectrum give information on the DM candidate.
This detection channel does not only give additional
information on the new particles, but it might as well be the
only kinematically allowed process at the ILC.
This is the case when only one new particle
exists in the energy range of the ILC.
In supersymmetric scenarios the other SUSY masses might
be just above the production threshold, preventing
the production of the LSP via the decay of heavier particles.
Without the need for heavier particles to be produced, ISR searches
extend the ILC reach on WIMP masses to $M_\chi \leq 250$~GeV,
just below half the CMS energy.
A model independent description of ISR associated
WIMP pair production can be derived as in~\cite{Birkedal:2004xn}.
Under the assumption of only one new particle responsible for
the CDM the annihilation cross section of a WIMP pair
into an $e^+e^-$ pair can be estimated from
the observed DM relic density~(Fig.~\ref{Fig:Birk}(a)).
Exploitation of crossing symmetry yields the cross section for
the inverse process~(Fig.~\ref{Fig:Birk}(b)). For
soft or collinear photons, factorization theorems
can be applied to get the observable signal process depicted in
Figure~\ref{Fig:Birk}(c). It can be shown
that the soft/collinear approximation holds also for 
photons with larger polar angles~\cite{Birkedal:2004xn}.
The resulting differential production cross section as a function of the emitted photon energy
is given by:
\begin{eqnarray}
\frac{d\sigma}{dx}\sim \kappa_e(P_e,P_p)
2^{2J_0}(2S_{\chi}+1)^2 \left(1-\frac{4M_{\chi}^2}{(1-x)s}\right)^{1/2+J_0}.
\end{eqnarray}\label{Eq:sigma}
where $x = 2E_{\gamma}/\sqrt{s}$ and $\kappa_e(P_e,P_p)$
is the polarization dependent squared coupling to the electrons 
in the initial state. The parameter $J_0$ is the angular momentum
of the dominant partial wave in the production process.
Depending on the nature of the interaction $J_0$ is related
to the spin of the exchange particle. Since all known models
predict either s-wave~($J_0 = 0$) and p-wave~($J_0 = 1$) WIMP annihilation
only these cases are considered here.
\begin{figure}[htb]
\setlength{\unitlength}{1cm}
\begin{picture}(12.0,5.0)
\put(0.0,0.0){\epsfig{file=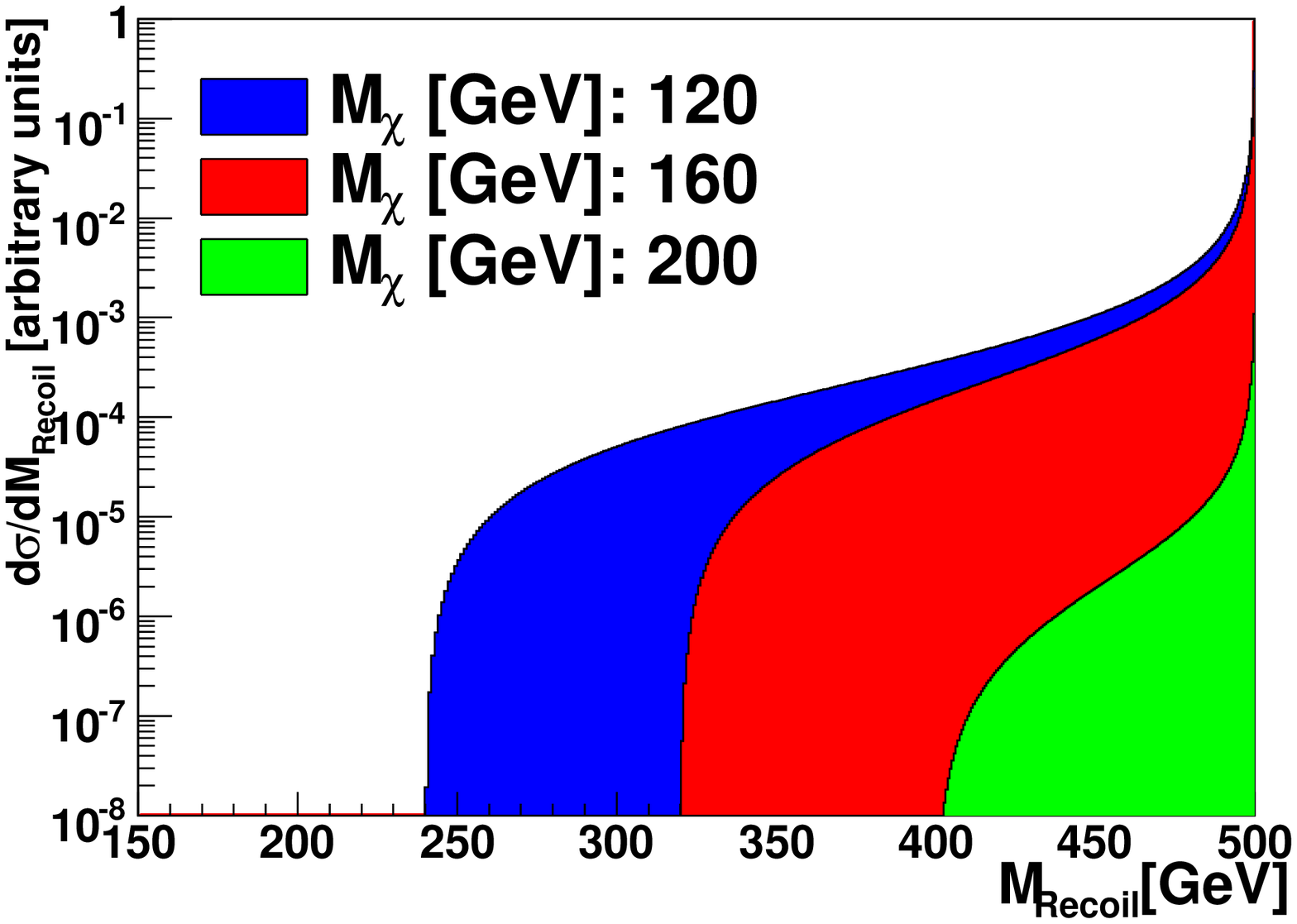,scale=.33}}
\put(7.5,0.0){\epsfig{file=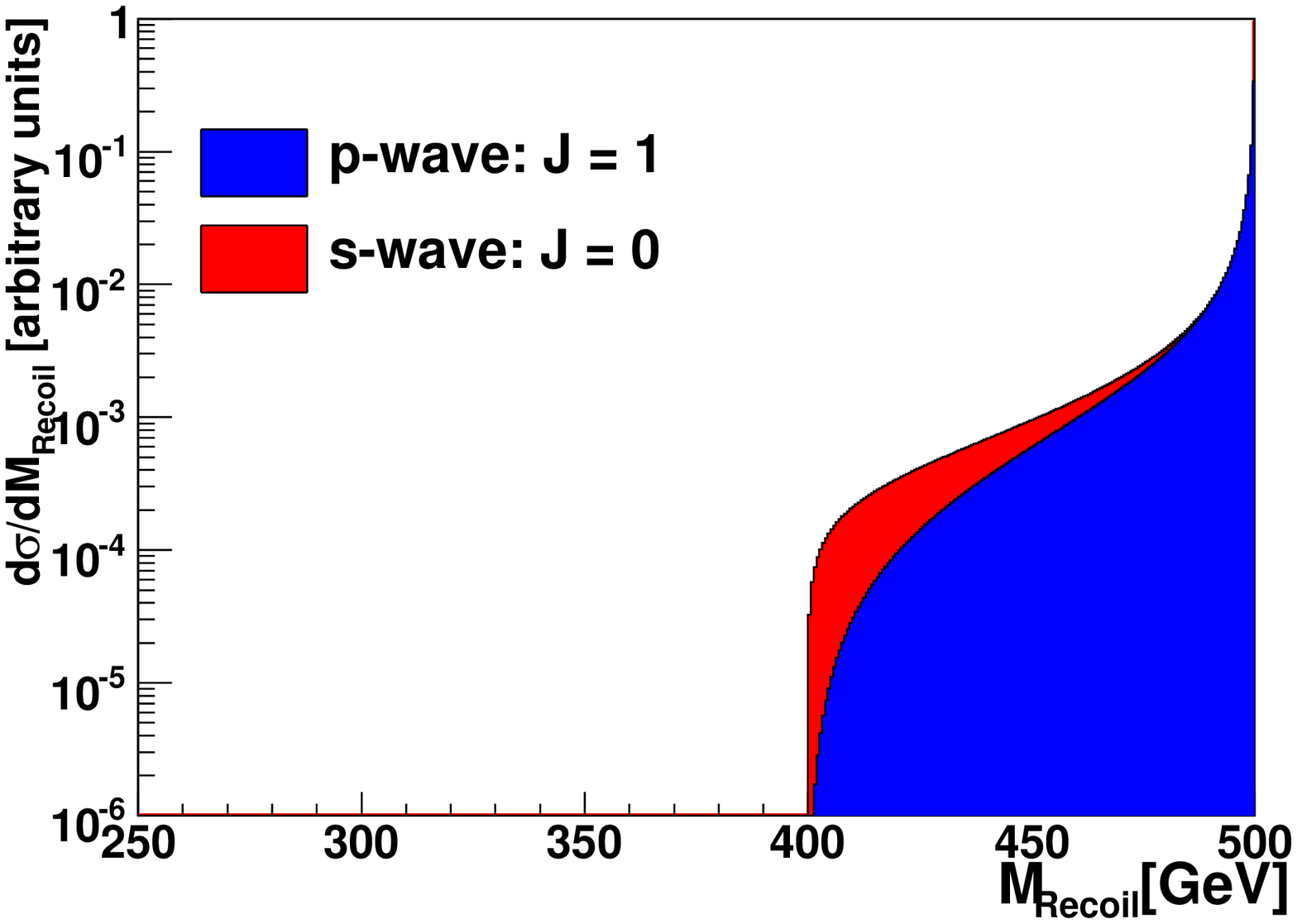,scale=.33}}
\put(0.0,0.0){(a)}
\put(7.5,0.0){(b)}
\end{picture}
\caption{Differential $\chi\chi\gamma$ production cross section in terms of 
the recoil mass for~(a) different WIMP masses and~(b) 
$J_0 =0$ and $J_0 =1$~(in arbitrary units).}\label{Fig:Recoil}
\end{figure}
Figure~\ref{Fig:Recoil} show the cross section as function of the recoil mass
$M_{recoil}^2 = s-2\sqrt{s}E_{\gamma}$ for (a) WIMP masses of 120, 160 and 200~GeV and
(b) for a 200~GeV WIMP with $J_0 = 0$ and $J_0 = 1$. The cut-off in the spectrum
is related to the WIMP mass and the shape at the threshold provides information
on the partial wave quantum number. In order to measure the energy cut-off and
the shape, a very good energy resolution for
photon detection is demanded from the detector and the reconstruction
algorithms.\\
This analysis is performed with a full simulation
of the International Large Detector concept~(ILD)~\cite{:2010zzd}.
A quadrant of the ILD is shown in Figure~\ref{Fig:Quad}.
The ILC's capability of delivering polarized beams can be
used to significantly suppress SM backgrounds, and
increase possible signals depending of
the structure of the involved interactions.
The role of beam polarization at the ILC is discussed in~\cite{MoortgatPick:2005cw}.

\section{ILD detector simulation and photon reconstruction}
\begin{wrapfigure}{r}{0.5\columnwidth}
\centerline{\includegraphics[width=0.45\columnwidth]{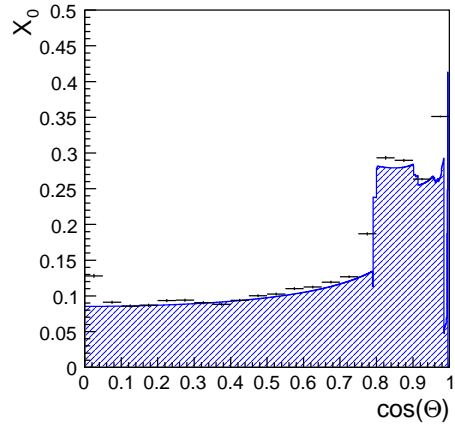}}
\caption{The material budget of the {\sc Mokka} implementation of
the {\tt ILD00} detector model tracking system in terms of the radiation length
$X_0$ as a function of the polar angle $\Theta$. For comparison the
budget as described in the LoI~\cite{:2010zzd} is plotted in the blue histogram.}\label{Fig:Rad}
\end{wrapfigure}
Single photon events provide the means for a fast 
validation of the detector model's implementation in the simulation framework.
Some key parameters like 
the overall material budget and the intrinsic energy resolution
of the calorimetric system for electromagnetic processes
can be determined from the analysis of
a large simulated sample of photons shot randomly into the detector.
Secondly some aspects of the employed reconstruction 
algorithms can be studied using single photon 
events, for example the quality
of pattern recognition for electromagnetic clusters, addressed
in Subsec.~\ref{Sec:rec}.

\subsection{Material budget}
A good understanding of the material budget
is important for the evaluation of the systematic effects from photon conversions on the
event reconstruction.  
The angular dependence of the material budget that
particles originating from the interaction point are exposed to on their
way out through the tracking subsystems~(vertex detector, TPC,
intermediate trackers and support structures) can be evaluated in the
detector simulation by counting the number $N_{gen}$ of photons 
generated and the number $N_{unconv.}$ of photons not converted 
before reaching the ECAL face.
The required information is given in the output of the
{\sc Mokka} implementation of the {\tt ILD00}.
The material budget in terms of the radiation length is given by:
\begin{eqnarray}
\frac{x}{X_0}   &  =  &  -\frac{9}{7} \ln{\left(\frac{N_{unconv.}}{N_{gen}}\right)}
\end{eqnarray}
Figure~\ref{Fig:Rad} shows the amount of material determined 
for the {\tt ILD00} detector model implemented in {\sc Mokka}~v-06-07-p01
The results are based on $10^{6}$ single photon 
events from the SM $\nu\nu\gamma$ process.
For reference the material budget for the tracking system
as quoted in the ILD LoI~\cite{:2010zzd}
is shown in the underlying histogram. The two most
visible features are the increased amount material
at $\cos{\Theta} \simeq 0.0$ which is due
to the dividing central cathode in the TPC,
and the additional material $\cos{\Theta} \ge 0.8$
in the forward region is related to the TPC
endplate.
  
\subsection{Photon reconstruction}\label{Sec:rec}
\begin{figure}[htb]
\setlength{\unitlength}{1cm}
\begin{picture}(13,7.0)
\put(0.0,0.0){\epsfig{file=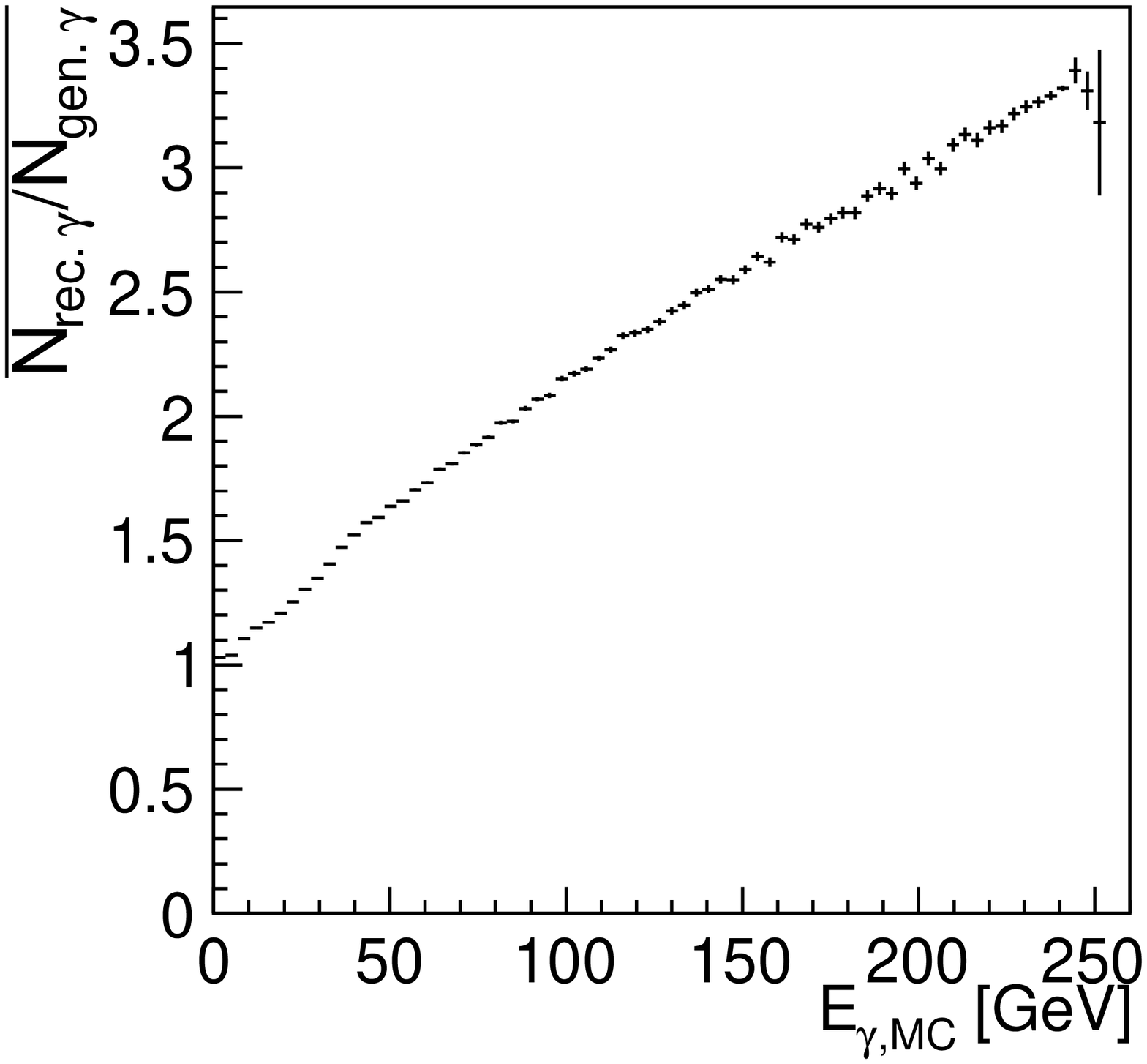,scale=.33}}
\put(7.5,0.0){\epsfig{file=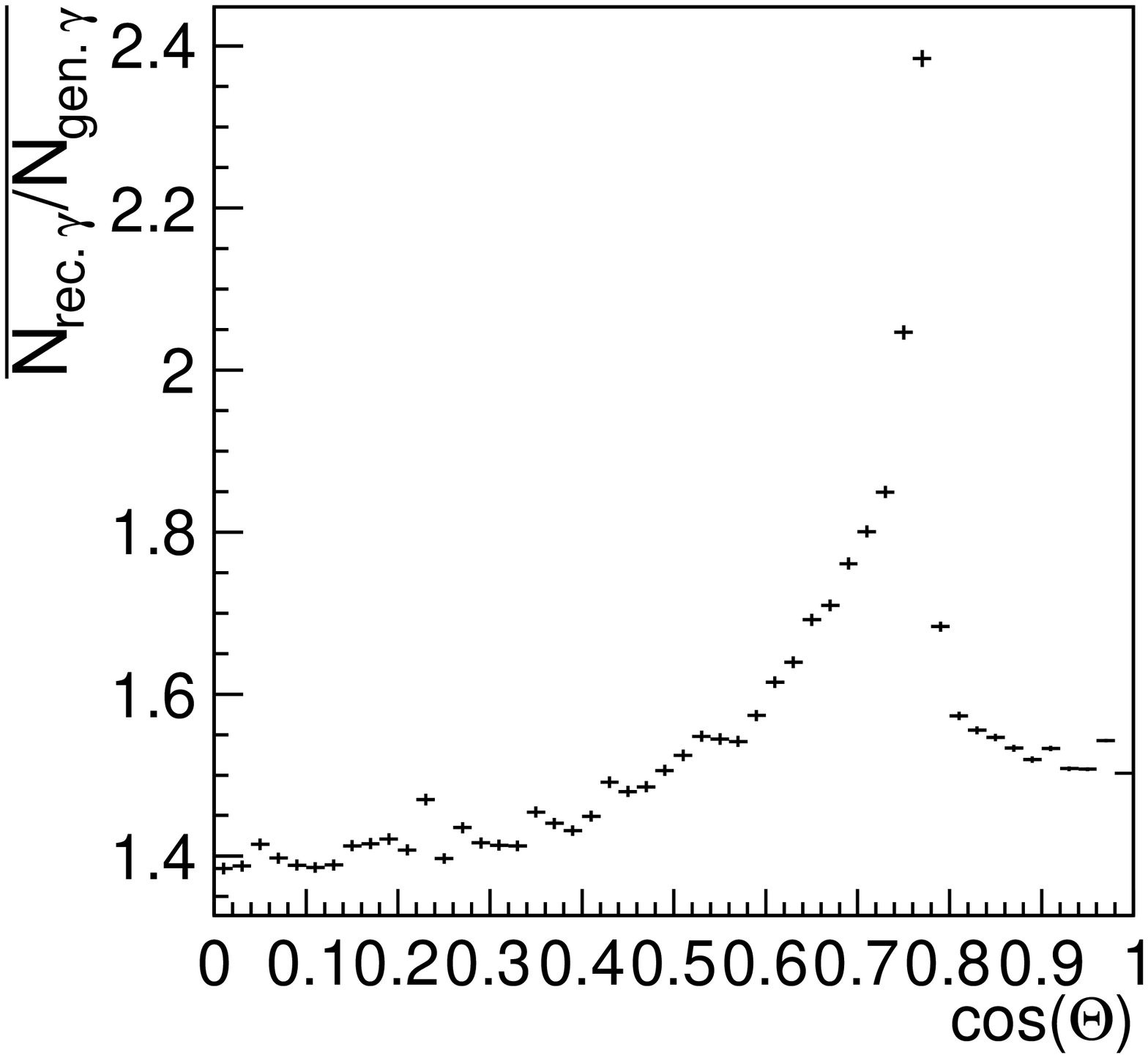,scale=.33}}
\put(0.0,0.0){(a)}
\put(7.5,0.0){(b)}
\end{picture}\caption{ Performance of the photon reconstruction in terms of the ratio
$\overline{N_{rec}/N_{gen}}$  of reconstructed per generated photons as function of photon energy~(a) and polar angle~(b).}  
\label{Fig:Receff1}
\end{figure}

\begin{figure}[htb]
\setlength{\unitlength}{1cm}
\begin{picture}(13,7.0)
\put(0.0,0.0){\epsfig{file=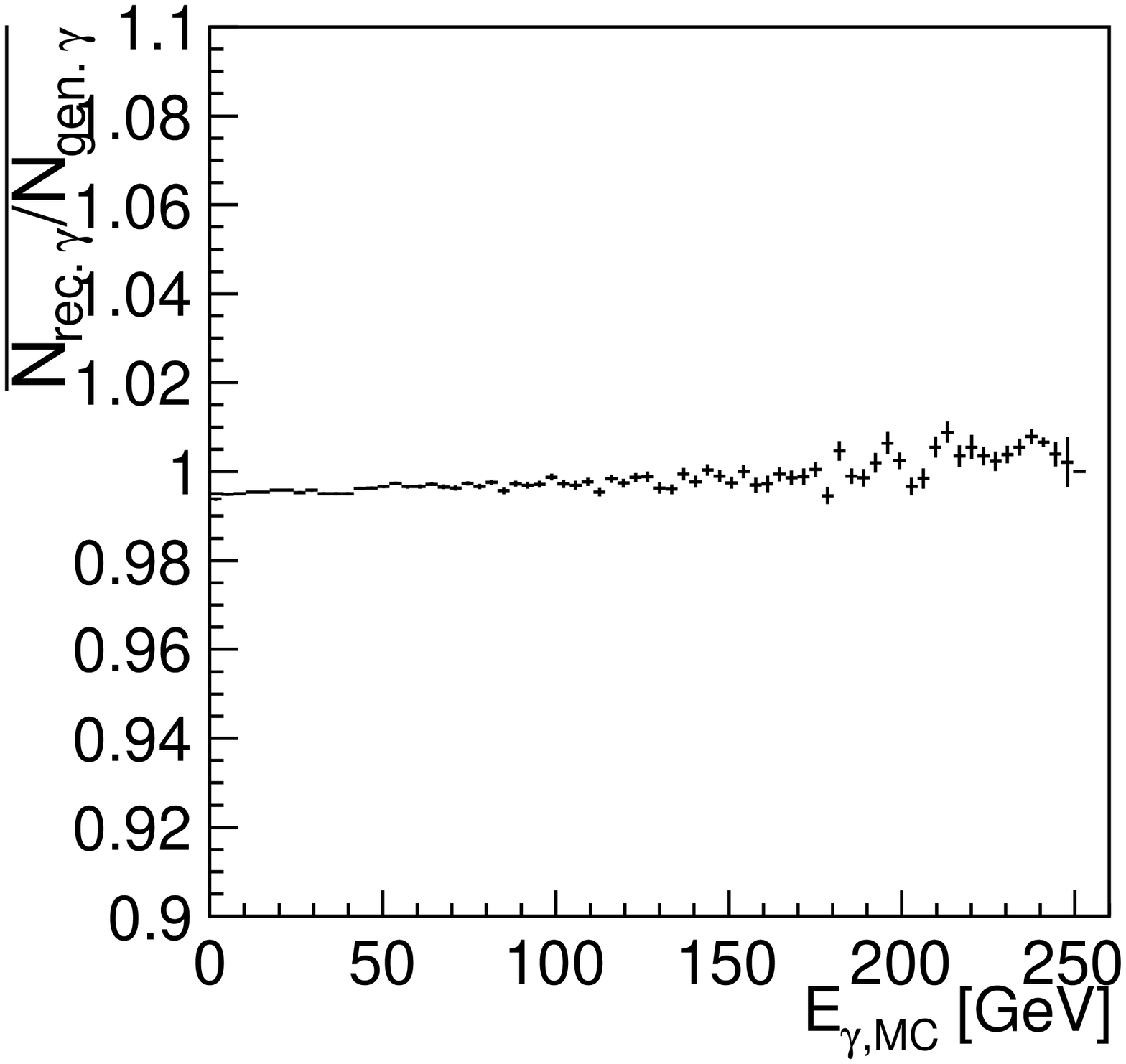,scale=.33}}
\put(7.5,0.0){\epsfig{file=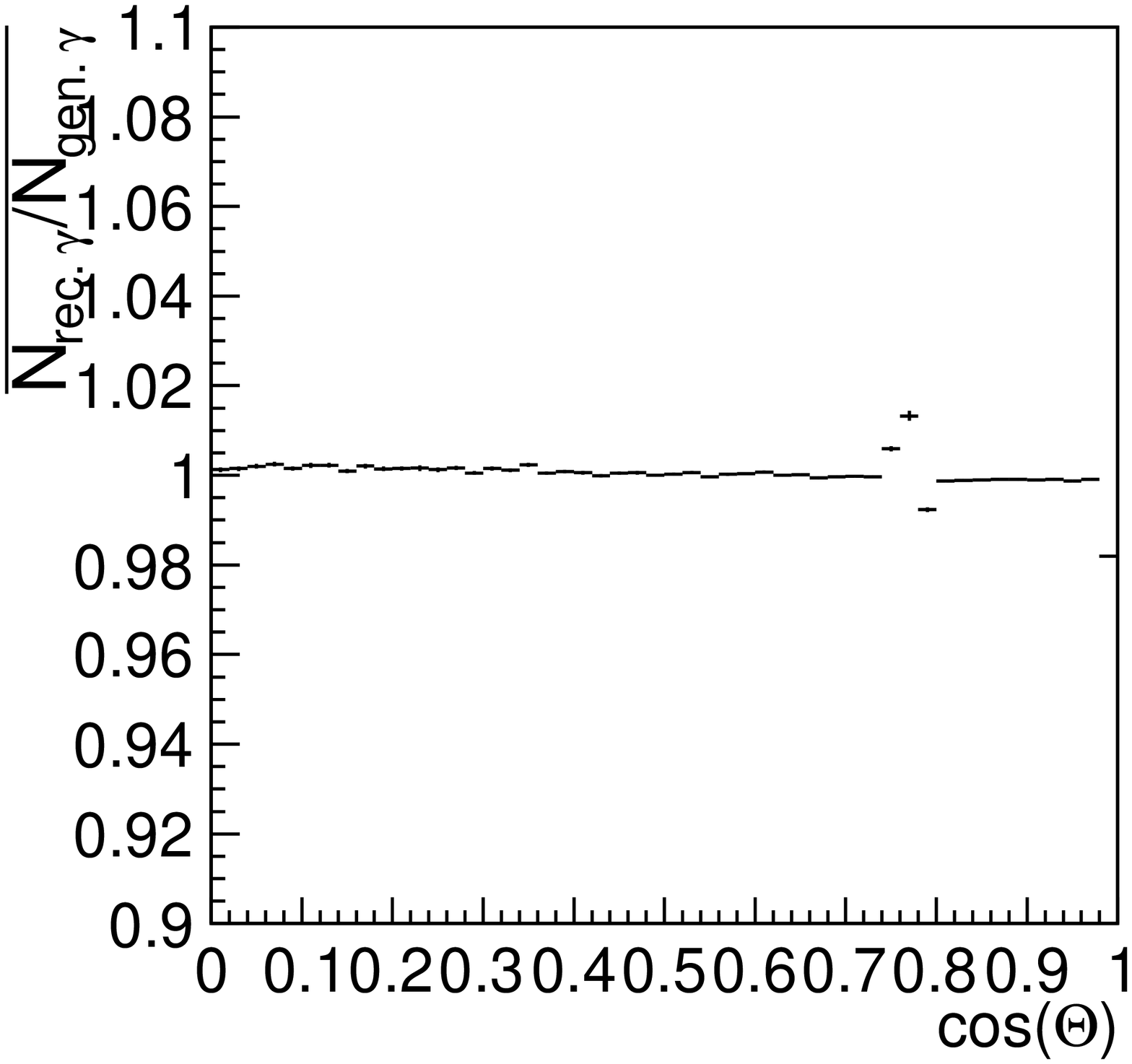,scale=.33}}
\put(0.0,0.0){(a)}
\put(7.5,0.0){(b)}
\end{picture}\caption{Ratio of $\overline{N_{rec}/N_{gen}}$ for photons after merging of
fractured photon objects.}\label{Fig:Receff2}
\end{figure}

The {\sc PFlow} paradigm requires that every particle
partaking in an event is to be reconstructed individually.
For single photon searches especially the performance of the
reconstruction algorithms in finding electromagnetic clusters 
in the ECAL is required to be very good. Figure~\ref{Fig:Receff1}
depicts the average ratio $\overline{N_{rec}/N_{gen}}$ 
of reconstructed photon candidates
per generated photon as a function of the generated photon energy~(a)
and generated photon polar angle $\cos{(\Theta)}$~(b).
For event reconstruction the {\sc Marlin} framework~v00-10-04 and
the {\sc PandoraPFlow}~\cite{Thomson:2009rp} algorithm~v-03-01 have been used. 
A photon is considered as reconstructed when a {\sc PFlow}
photon object is found within 0.1~rad~(seen from the IP) from a 
generated one. Additionally the photon object is required to be 
related to the generated photon on MC level.
The ratio continuously rises from one at low  photon energies 
to about three at $E_\gamma = 250$~GeV. The angular dependence
displays a steep rise at polar angles of $\cos{\Theta} \simeq 0.8$. 
This angular region corresponds to the insensitive region between 
the ECAL barrel and ECAL endcap where pattern recognition is difficult
due to the complex geometry~(see Fig.~\ref{Fig:Quad}). 
For the analysis presented in
the next section this effect
is countered by combining the ``split'' photon clusters
into higher level photon objects. Photon clusters that
are less apart than 0.04~rad are considered as being fractured
remnants from a single incident photon. The result of the merging procedure
is shown in Figure~\ref{Fig:Receff2}, where the ratio
deviates from unity on the sub-percentage level.

\section{Determination of WIMP parameters}
WIMP production with associated ISR leads to an excess
in the photon spectrum at energies below 
\begin{equation}
E_\gamma^{max} = \frac{E_b^2-M_\chi^2}{E_b}.
\label{Eq:lim}
\end{equation}
The cross sections are in the order of a few fb, depending
on the values of the model parameters $\kappa_e(P_e,P_p)$, $J_0$, $M_\chi$ and
the WIMP spin $S_\chi$, resulting in an $S/B$ ratio of $10^{-2}--10^{-3}$ 
to the irreducible $\nu\nu\gamma$ background. The parameter
$\kappa_e(P_e,P_p)$ can be deconstructed according to the four
possible $e^+e^-$ helicity configurations:
\begin{eqnarray}
\kappa_e(P_e,P_p) = \frac{1}{4}(1+P_{e})[(1+P_{p})\kappa(e_{R}p_{L})+(1-P_{e^{+}})\kappa(e_{R}p_{R}
)]&&\nonumber\\
+\frac{1}{4}(1-P_{e})[(1+P_{p})\kappa(e_{L}p_{L})+(1-P_{e^{+}})\kappa(e_{L}p_{R})]
\end{eqnarray}
The WIMPs 
might couple only to right-handed electrons and left-handed positrons
($\kappa(e_R,p_L)$, helicity conserving). The couplings might in addition
also conserve parity~($\kappa(e_R,p_L) = \kappa(e_L,p_R)$). In both
cases the $S/B$ ratio can be enhanced using polarized beams.
The results presented here are based on the full SM
$\nu\nu\gamma(N)\gamma$ ($N = 0,1,2$) background simulated for the
{\tt ILD00} detector model implemented in {\sc Mokka}~v-06-07-p01.
For event generation the {\sc Whizard} event generator~\cite{Kilian:2007gr}
is used.
Event reconstruction is done within the {\sc Marlin} framework~v00-10-04 using
the {\sc PandoraPFlow} algorithm~v-03-01. Each event contains
up to two additional ISR photons from the electron and positron. 
In total an integrated luminosity of $\mathcal{L} = 250$ fb$^{-1}$
is generated and simulated. The signal itself is not generated
and simulated, but is obtained by assignment of weights to
the dominant $\nu\nu\gamma$ background, where the
weights are given by the ratio of the cross sections
for neutrino and WIMP pair production, $w=\sigma_{\chi\chi\gamma}/\sigma_{\nu\nu\gamma}$
evaluated for $E_\gamma$ and $\Theta$ of the
detected ISR photon.
For the signal to be statistically independent of the 
background only half of the simulated SM background is used
for the weighted signal, while the other half serves as
signal free background. An additional weight
is assigned to each event to adjust for the
demanded luminosity. The event selection requires
a single photon in the detector and no activity in the
tracking system.

\subsection{2$\sigma$ reach on $\kappa$}
\begin{figure}[htb]
\setlength{\unitlength}{1cm}
\begin{picture}(12.0,7.0)
\put(0.0,0.0){\epsfig{file=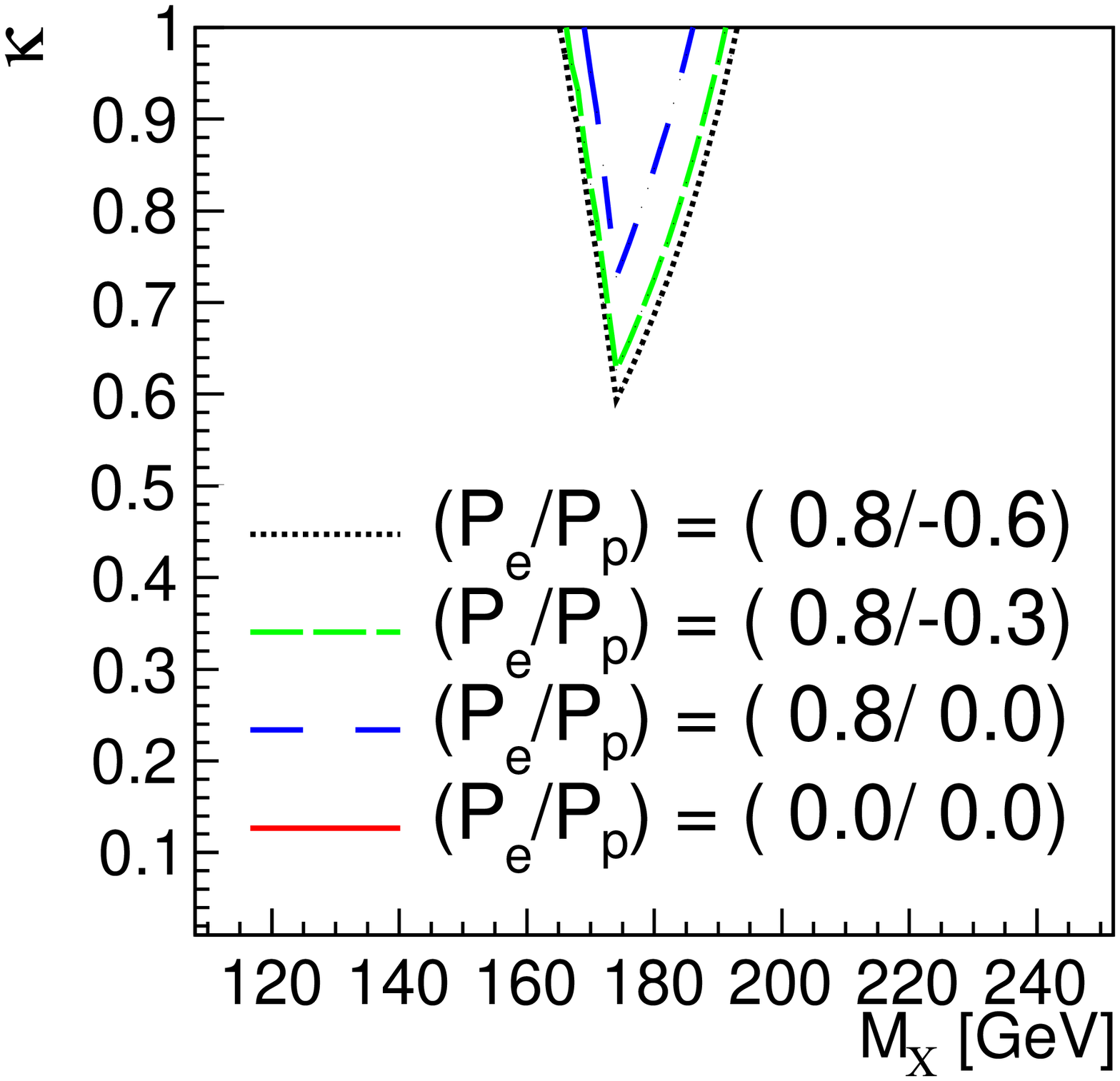,scale=.33}}
\put(7.5,0.0){\epsfig{file=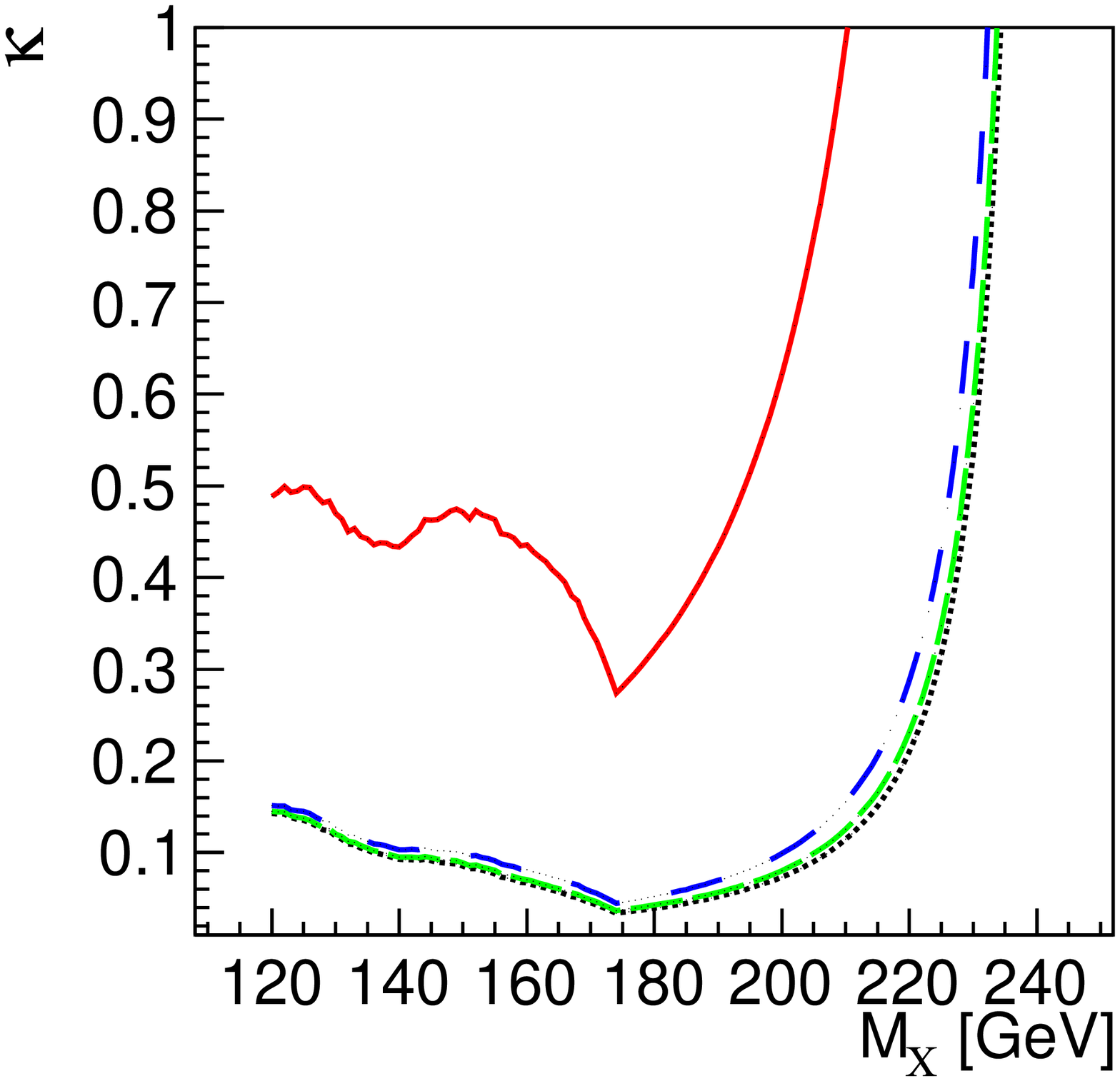,scale=.33}}
\put(0.0,0.0){(a)}
\put(7.5,0.0){(b)}
\end{picture}
\caption{$2\sigma$ reach for 200~fb$^{-1}$ on $\kappa_e(P_e,P_p)$ as
function of the WIMP mass. The results are given for (a) spin 0 WIMPs with
helicity and parity conserving couplings and (b) spin 1 WIMPs with couplings
conserving only helicity. The contour lines are given for several beam polarizations,
see text.}\label{Fig:K1}
\end{figure}

The reach on 2$\sigma$ level of the ILC on the
parameter $\kappa$ as function of the WIMP mass
is shown in Figure~\ref{Fig:K1}(a). The plot corresponds
to an integrated luminosity of 200 fb$^{-1}$.  The WIMP parameters 
are chosen as $S=0$, $J_0 = 1$ and the WIMP couplings to electrons
conserve helicity and parity. The contour lines give
the reach on the parameter $\kappa$ as a function
of the WIMP mass.  The area above the curves is accessible.
For each WIMP mass a signal region is defined in the energy distribution of
the emitted photons, $E_{min}\le E_\gamma\le E_{max}$. 
The lower cut on the photon energies 
ensures that the recoiling WIMP system is non-relativistic.
This has to be enforced since only s-wave and p-wave production is assumed.
The upper cut is given by the kinematic limit~(Eq.~\ref{Eq:lim}).
This cut serves to increase the signal to background ratio.  
The result is plotted for different beam polarizations:
\begin{itemize}
 \item $(P_e/P_p) = (0.0/0.0)$ (red, solid)
 \item $(P_e/P_p) = (0.8/0.0)$ (blue, wide dashed)
 \item $(P_e/P_p) = (0.8/-0.3)$ (green, dashed)
 \item $(P_e/P_p) = (0.8/-0.6)$ (black, dotted)
\end{itemize}
In this scenario and the given luminosity,
no value of $\kappa$ is accessible for unpolarized beams, hence no 
WIMP signal can be distinguished from the SM background
on $2\sigma$~level~(the corresponding contour line coincides
with $\kappa =1$).
With increasing beam polarization
the reach is extended for WIMP masses between 160 and 200~GeV down
to values of 0.6. The loss of sensitivity to higher masses
is due to the kinematic limit, while for lower masses the
signal is increasingly reduced by the lower cut on the photon energies.
Figure~\ref{Fig:K1}(b) shows the reach for the
case where the WIMPs~($S=1$, $J_0 = 1$), couple only to right handed electrons
and left handed positrons. Even for unpolarized beams the reach covers
large areas of the $M-\kappa$ plane for masses up to 200~GeV.
With increasing beam polarizations the SM background is suppressed and
the signal enhanced, resulting in a large increase of accessible
values down to $\kappa \le 0.1$.

\subsection{Mass measurement}
\begin{figure}[htb]
\centering
\includegraphics[width=0.6\linewidth]{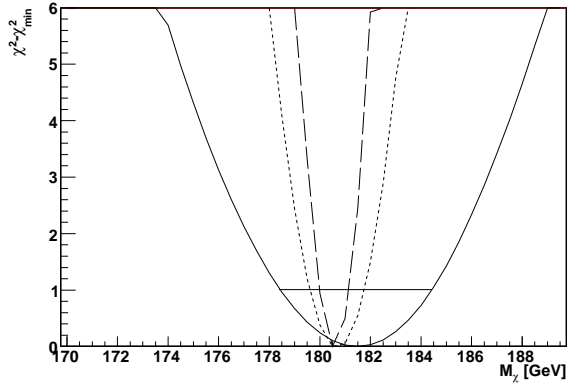}
\caption{$\chi^2$ of comparison of 180~GeV spin 1 WIMP ($\kappa = 0.3$,$J_0 =1$) signal distribution with
signal mass templates for three different polarization configurations, as explained in
the text.
}\label{Fig:K3}
\end{figure}

This section on the WIMP mass measurement refers to
results from 2007~\cite{Bartels:2009fa}. The simulated detector model
is {\tt LDCPrime\_02Sc}.
The mass of a WIMP candidate can be determined by comparing the signal 
spectrum to template spectra of different WIMP mass hypotheses but with
otherwise identical parameters.
Figure~\ref{Fig:K3}(a) shows the $\chi^2$ for
the template fits as a function of the template mass.
The signal is a 180~GeV spin 1 WIMP with couplings conserving 
helicity and parity. The other parameters are set to
$\kappa = 0.3$ and $J_0 =1$. The simulated data is equivalent
to an integrated luminosity of 500 fb$^{-1}$.
Three different polarization configurations are considered:
\begin{itemize}
 \item $(P_e/P_p) = (0.0/0.0)$ (solid)
 \item $(P_e/P_p) = (0.8/0.0)$ (dotted)
 \item $(P_e/P_p) = (0.8/-0.6)$ (dashed)
\end{itemize}
Depending on the beam polarization the mass can be determined to:
\begin{itemize}
 \item $M = 181.5 \pm 3.0$~GeV ($(P_e/P_p) = (0.0/0.0)$)
 \item $M = 180.5 \pm 0.5$~GeV ($(P_e/P_p) = (0.8/-0.6)$)
\end{itemize}
In this case due to the preferable choice of the WIMP couplings
highly polarized beams increase the mass resolution by a factor of six.

\subsection{Simultaneous determination of mass and $J_0$}
As pointed out in Section~\ref{Sec:2} the shape of the
photon distribution at the threshold provides information on the
dominant partial wave of the production process.
The energy distribution of ISR photons from
the SM background and the WIMP production can be fitted
with the WIMP cross section of formula~\ref{Eq:sigma},
leaving the WIMP mass and the partial wave $J_0$ 
as free parameters in the fit. For $\mathcal{L} = 100$~fb$^{-1}$ ($(P_e/P_p) = (0.8/-0.6)$)
this method recovers a WIMP mass of $M\chi= 173.8\pm5.1$~GeV
and $J_0 = 1.26\pm 0.34$ for input parameters of
 $M\chi= 180$~GeV and $J_0 = 1$ respectively. The remaining input parameters are
$\kappa = 1.0$, $S =1$ and the couplings conserve helicity only.
Although the luminosity is low compared
with the ILC design, the s-wave production can be excluded on
3$\sigma$ level, due to the assumed WIMP couplings and
beam polarization. The simultaneous mass fit performs
significantly worse than the dedicated template fit
of the previous section. Since the mass can be determined 
independently from $J_0$, a constraint could be imposed on
it in the simultaneous fit. The expected improvement
on the  $J_0$ measurement has to be quantified.     

\section{Conclusions}
The study presented here shows that there is a good chance of a 
model independent detection of WIMPs at the ILC. If the
WIMP couplings to electrons are not too small, their
masses can be measured on the percentage level. Both the
detection sensitivity and the mass resolution is
enhanced when polarized beams are assumed. First results
indicate that even the more involved measurement of
the partial wave quantum number seems to be possible,
given a good understanding of the SM background. 

\section{Acknowledgments}
The authors acknowledge the support by DFG grant Li 1560/1-1.

\section{Bibliography}

% ****************************************************************************
% BIBLIOGRAPHY AREA
% ****************************************************************************

\begin{footnotesize}
% IF YOU DO NOT USE BIBTEX, USE THE FOLLOWING SAMPLE SCHEME FOR THE REFERENCES
% ----------------------------------------------------------------------------

% ----------------------------------------------------------------------------

\end{footnotesize}

% ****************************************************************************
% END OF BIBLIOGRAPHY AREA
% ****************************************************************************

\end{document}